\documentclass[aps,prl,twocolumn,floatfix,superscriptaddress,nobibnotes,10pt,hidelinks,nofootinbib]{revtex4}

\usepackage{amsmath}
\usepackage{amssymb}
\usepackage{braket}
\usepackage{graphicx}
\usepackage{hyperref}
\usepackage[version=4]{mhchem}
\usepackage[separate-uncertainty=true,multi-part-units=single]{siunitx}
\begin{document}

\title{Synthetic spin-orbit coupling in superconductor-semiconductor hybrid nanowires with micromagnet arrays}

\author{M.P.~Hynes}
\thanks{Present Address: C12 Quantum Electronics, Paris, France}
\affiliation{London Centre for Nanotechnology, University College London, London WC1H 0AH, United Kingdom}
\affiliation{Department of Physics and Astronomy, University College London, London, WC1E 6BT, United Kingdom}

\author{D.~Burke}
\affiliation{Blackett Laboratory, Imperial College London, London, SW7 2AZ, United Kingdom}

\author{K.~Ganesh}
\affiliation{Blackett Laboratory, Imperial College London, London, SW7 2AZ, United Kingdom}

\author{A.~Vekris}
\affiliation{Center for Quantum Devices, Niels Bohr Institute, University of Copenhagen, Universitetsparken 5, 2100 Copenhagen \O, Denmark}

\author{B.J.~Villis}
\affiliation{London Centre for Nanotechnology, University College London, London WC1H 0AH, United Kingdom}

\author{J.C.~Gartside}
\affiliation{Blackett Laboratory, Imperial College London, London, SW7 2AZ, United Kingdom}

\author{T.~Kanne}
\affiliation{Center for Quantum Devices, Niels Bohr Institute, University of Copenhagen, Universitetsparken 5, 2100 Copenhagen \O, Denmark}

\author{J.~Nyg{\aa}rd}
\affiliation{Center for Quantum Devices, Niels Bohr Institute, University of Copenhagen, Universitetsparken 5, 2100 Copenhagen \O, Denmark}

\author{K.~Moors}
\thanks{Present Address: Imec, Kapeldreef 75, 3001 Leuven, Belgium}
\affiliation{Peter Gr\"unberg Institut (PGI-9), Forschungszentrum J\"ulich, 52425 J\"ulich, Germany}
\affiliation{JARA-Fundamentals of Future Information Technology, J\"ulich-Aachen Research Alliance, Forschungszentrum J\"ulich and RWTH Aachen University, 52425 J\"ulich, Germany}

\author{W.R.~Branford}
\affiliation{Blackett Laboratory, Imperial College London, London, SW7 2AZ, United Kingdom}

\author{M.R.~Connolly}
\affiliation{Blackett Laboratory, Imperial College London, London, SW7 2AZ, United Kingdom}

\author{M.R.~Buitelaar}
\affiliation{London Centre for Nanotechnology, University College London, London WC1H 0AH, United Kingdom}
\affiliation{Department of Physics and Astronomy, University College London, London, WC1E 6BT, United Kingdom}

\begin{abstract}
\vspace{0.2cm}
Spin-orbit interaction accounts for the coupling of momentum and spin degrees of freedom of electrons and holes in semiconductor materials. In quantum information processing, it allows for electrical control of spin states and for the engineering of topologically protected Majorana zero modes. Although such functionalities were previously considered to be limited to semiconductor materials with strong intrinsic spin-orbit interactions only, recent theoretical work proposes using external rotating magnetic fields to engineer synthetic spin-orbit coupling. This would relax material constraints and open up new research directions for materials with low intrinsic spin-orbit interaction or augment existing spin-orbit interaction in materials in which this interaction is already strong. Here we demonstrate the feasibility of this approach and introduce rotating magnetic fields along an InAs/Al hybrid nanowire using permalloy micromagnet arrays which yields an estimated synthetic Rashba spin-orbit interaction coefficient of 0.022 eV nm. We use transport spectroscopy and the energy dependence of Andreev bound states in the nanowires as a probe of the magnetic field profiles of the micromagnets which are reconfigurably prepared in parallel or antiparallel magnetization configurations.
\end{abstract}

\date{\today}

\maketitle

\section{Introduction}

The ability to control charge, spin, and orbital degrees of freedom at the nanoscale is at the basis of the considerable progress made in fields ranging from spintronics to quantum information processing. An example is the demonstration of spin-orbit qubits in III-V semiconductor materials such as InAs nanowires ~\cite{Nadjperge2010, petersson2012, wang2018,ungerer2024}. The strong intrinsic Rashba spin-orbit interaction in InAs couples momentum and spin degrees of freedom such that local electric fields can be used to coherently control the electron spin states. Further possibilities arise when Al is used to epitaxially cover semiconductor nanowires such as InAs or InSb. In these hybrid superconductor-semiconductor materials, the Al superconductor couples electron and hole states, giving rise to spin-orbit split Andreev bound state (ABS) levels and spin-dependent supercurrents~\cite{hays2021, pitavidal2023}. Superconductor-semiconductor materials are also intensively used in efforts to engineer p-wave pairing from s-wave superconductors by exploiting the interplay of spin-orbit interaction, superconductivity, and magnetism and realise topologically protected Majorana qubits~\cite{oreg2010,sau2010,das2012,vanheck2016,lutchyn2018,prada2020, microsoft2025}.

\begin{figure*}
    \centering
    \includegraphics[width=178mm]{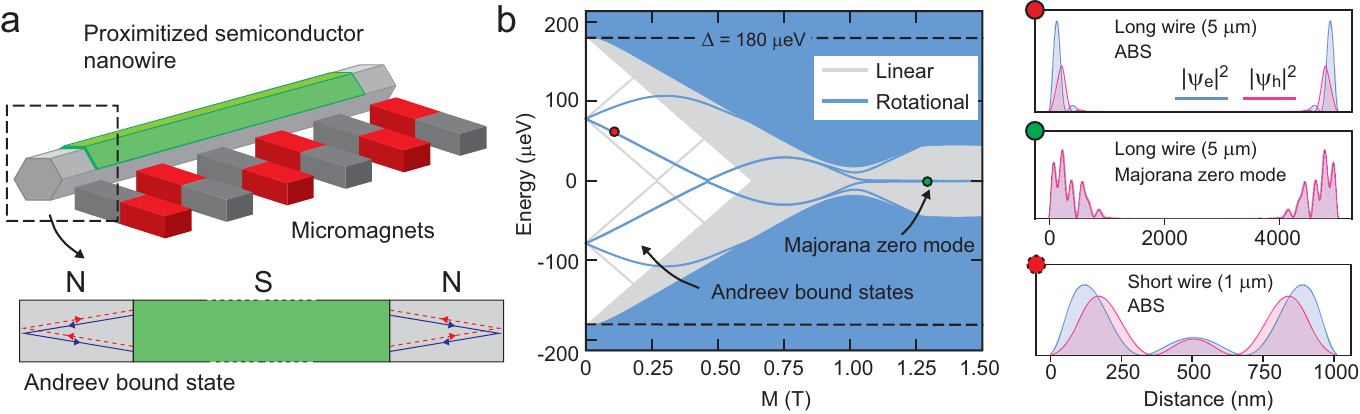}
    \caption[Synthetic spin-orbit coupling: Andreev bound states and Majorana zero modes]{
    \textbf{Synthetic spin-orbit coupling: Andreev bound states and Majorana zero modes.}
    \textbf{(a)} Schematic of a proximitized semiconductor nanowire with normal ends coupled to a micromagnet array which induces a rotating magnetic field texture. Andreev bound states form at the nanowire ends due to Andreev reflection at the normal-superconductor boundary in which an incident electron (blue lines) forms a Cooper pair in the superconductor region with the retroreflection of a hole of opposite spin and velocity in the normal region (red dashed lines). \textbf{(b)} Tight-binding simulations of the excitation spectrum of a hybrid superconductor-semiconductor nanowire of length 5 $\mu$m coupled to a magnetic stray field texture $M$. Blue and grey lines show the spectra for rotating and linear fields, respectively. For small magnetic fields, the spectra are dominated by Andreev bound states. For large magnetic fields - and for the rotational texture only - Majorana zero modes are observed in the calculations. The right plots show the electron $\lvert\psi_e\rvert^2$ and hole $\lvert\psi_h\rvert^2$ probability densities for the different eigenstates (coloured dots). For small magnetic fields and long nanowire lengths, Andreev bound states are well separated at the nanowires ends. For shorter nanowire lengths (bottom right plot), the wavefunction overlap pushes the antibonding state into the continuum above the superconducting gap, leaving only a single observable Andreev bound state in the excitation spectrum.
    }
    \label{fig:1}
\end{figure*}

Notwithstanding the progress made, significant challenges remain in material quality and device functionality for quantum information processing using superconductor-semiconductor materials. The demonstration of Majorana qubits in semiconductor materials is particularly challenging as in the presence of disorder, Andreev bound states readily mimic signatures of Majorana zero modes~\cite{prada2020,dassarma2021} and fault-tolerant braiding and fusion operations have not yet been demonstrated \cite{microsoft2025}.

It is in this context that recent proposals suggest overcoming material and device challenges by synthesising spin-orbit fields at the microscopic level by engineering rotating magnetic fields along a nanowire using micromagnet arrays as illustrated in Fig.~\ref{fig:1} \cite{braunecker2010,klinovaja2012a,kjaergaard2012,maurer2018,turcotte2020,kornich2020}. The main idea is that this allows engineering a Hamiltonian of the form:
\begin{equation}
    H=\Big(\frac{p_x^2}{2m}-\mu\Big)\tau_z+\frac{1}{2} g \mu_B \textbf{B}(x) \cdot \boldsymbol{\sigma} + \Delta \tau_x
\end{equation}

\noindent which consists of a kinetic term with effective electron mass $m$, chemical potential $\mu$, a magnetic field texture $\textbf{B}(x)$ along the nanowire induced by the micromagnets with $g$ the effective Land\'e g-factor, a proximity-induced superconducting pair potential $\Delta$ and where $\sigma_i$ and $\tau_i$ are Pauli matrices in spin and electron-hole space, respectively. Note that this Hamiltonian does not contain any (intrinsic) spin-orbit interaction. However, by choosing a rotating magnetic field texture such as: $\textbf{B}(x)=B_0 [\textrm{sin}(2 \pi x/R),\textrm{cos}(2 \pi x/R), 0]$, where $R$ is the spatial period of the field, the Hamiltonian of Eq.~1 maps onto \cite{braunecker2010,kjaergaard2012}:
\begin{equation}
    H=\Big(\frac{p_x^2}{2m}-\mu^* \Big)\tau_z+ \alpha_{\mathrm{eff}} p_x \sigma_y \tau_z+\frac{1}{2} g \mu_B B_0 \sigma_z + \Delta \tau_x
\end{equation}

\noindent where $\alpha_{\mathrm{eff}}=\hbar/2mR$ appears as an effective Rashba spin-orbit interaction coefficient and $B_0$ is the amplitude of an effective Zeeman field perpendicular to the spin-orbit direction. Here $\mu^*$ is the chemical potential, as in Eq.~1, but with a slight and constant offset \cite{kjaergaard2012}.

This approach offers several advantages. Firstly, it relaxes the material requirements and enables the exploration of low-disorder materials with low intrinsic spin-orbit interaction, opening up new directions to implement Majorana qubits for topological quantum computing~\cite{desjardins2019}. Secondly, in semiconductor nanowires that combine intrinsic and synthesised spin-orbit fields it has been predicted to give rise to a wide range of novel topological phases~\cite{klinovaja2012a,rainis2014,rex2020,marra2022,kobialka2021}. This would provide more robust Majorana bound states and remove the need for external magnetic fields \cite{jardine2021}. Thirdly, controlling the periodicity of the micromagnet array directly affects the synthetic spin-orbit coupling strength and, consequently, the size of the topological energy gap~\cite{sau2010a}.

This is further illustrated in Fig.~\ref{fig:1}b which shows the simulated energy spectrum of a hybrid superconductor-semiconductor nanowire as a function of micromagnet stray field magnitude $M$ coupled to the nanowire. In anticipation of results shown below, the simulations take into account both intrinsic spin-orbit interaction and synthetic spin-orbit interaction induced by micromagnets as well as both a proximity-induced superconducting central section of the nanowire and normal ends. Spectra are shown for both a linear field (grey lines) and a rotational field texture (blue lines), see Supplementary Section I for details. At low magnetic fields, the device is characterised by the presence of Andreev bound states that arise at the normal-superconducting interfaces at both ends of the nanowire as shown in Fig.~\ref{fig:1}a. For large magnetic field strengths - and the rotational field texture only - Majorana zero modes are observed which is also apparent from the complete overlap of the electron and hole wavefunction densities as shown in the rightmost plots of Fig.~\ref{fig:1}b (green dot). Importantly, the observation of the Majorana modes does not require any global external magnetic field.

\begin{figure*}
    \centering
    \includegraphics[width=178mm]{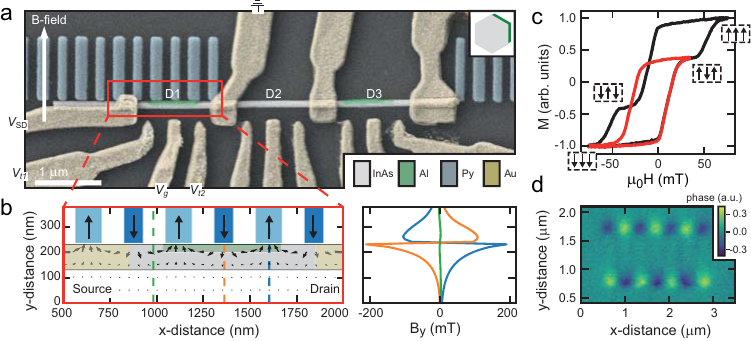}
    \caption[Nanowire quantum dot with micromagnet array.]{
    \textbf{InAs semiconductor nanowire device with micromagnet array.}
    \textbf{(a)} False colour scanning electron micrograph of the device consisting of InAs nanowire (light grey), Al half-shell regions (green), Ti/Au contacts and gates (yellow), and permalloy (NiFe) micromagnet arrays (blue). There are three sections along the single nanowire, two sections with superconductivity (D1 and D3) and a section without (D2). Section D1 has nearby micromagnets while D2 and D3 are control devices. The inset shows a schematic cross-section of the wire, including Al on two of the six side facets.
    \textbf{(b)} Left: simulated magnetic field profile generated by antialigned micromagnets. The micromagnet field rotates along the length of the nanowire as indicated by the arrows, decreasing in magnitude across the width of the wire. Right: $y$-component of the magnetic field generated by the micromagnets through the three different cuts indicated by the dashed lines in the left panel. \textbf{(c)} Magneto-optic Kerr effect (MOKE) data from many repeated micromagnet arrays with dimensions similar to those used in panel (a), showing magnetisation plateaus at two different critical magnetic fields, corresponding to the antiparallel and parallel micromagnetic configurations as indicated by the arrows. The major loop is shown in black, while the minor loop is shown in red.
    \textbf{(d)} Magnetic force micrograph of a micromagnet array showing an antiparallel magnetization configuration.
    }
    \label{fig:2}
\end{figure*}

Although therefore of considerable interest, superconductor-semiconductor devices coupled to micromagnets showing engineered rotating magnetic fields have not been demonstrated yet. In our work, we achieve this by addressing key experimental challenges such as precise nanowire positioning and the ability of in-situ micromagnet configuration control, to fabricate and characterise a superconductor-semiconductor InAs nanowire proximitized by an Al shell and coupled to a permalloy (NiFe) micromagnet array. We demonstrate the ability to switch individual micromagnets within the array and, using transport spectroscopy, show that the micromagnets are capable of generating a rotating magnetic field along the nanowires by preparing them in an antiparallel magnetization configuration. Taking advantage of the relatively large $g$-factors of InAs we measure differences in induced Zeeman fields of the order \qty{40}{\micro\electronvolt} between parallel and antiparallel magnet configurations, consistent with model calculations such as shown in Fig.~1. The induced Zeeman fields of the permalloy micromagnets used in our work are not strong enough to observe Majorana zero modes – given the parent Al superconducting gap energy $\Delta=180$ $\mu$eV. In our outlook we provide estimates of the engineered field strengths and material properties required to attain a topologically protected phase.

\section{Superconductor-semiconductor hybrid device}

\begin{figure*}
    \centering
    \includegraphics[width=178mm]{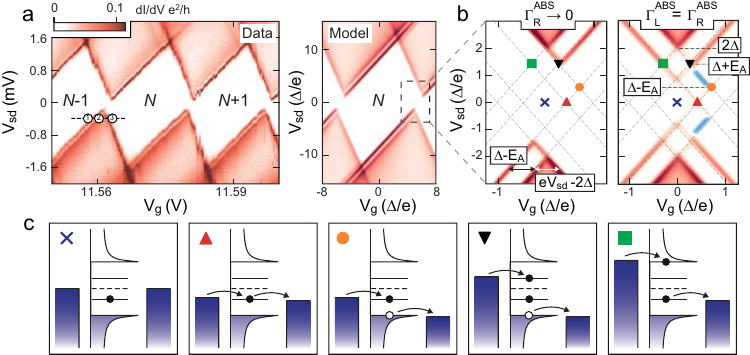}
    \caption[Different tunnelling regimes of the subgap state.]
    {\textbf{Transport spectroscopy and Andreev bound state model.} \textbf{(a)} Left: Measured differential conductance as a function of applied gate voltage $V_{\text{g}}$ and source-drain bias voltage $V_{\text{sd}}$ showing Coulomb diamonds with charging energy $\sim 1.5$ meV. The data are characterised by an absence of low-bias conductance and an excitation line along the side of the Coulomb diamonds with positive slope. Right: Corresponding model calculations for the hybrid quantum dot using a vanishing coupling between the ABS levels to the drain electrode $\Gamma_R^{\text{ABS}} \rightarrow 0$. \textbf{(b)} Left: Model calculation of the low-bias differential conductance for $\Gamma_R^{\text{ABS}} \rightarrow 0$. The various diagonal dashed lines indicate energies for which the Andreev bound states and the BCS continuum edge are aligned with the electrochemical potentials of the source and drain electrodes. The symbols indicate different scenarios shown in panel (c). Right: As the left panel but for finite symmetric coupling $\Gamma_L^{\text{ABS}}=\Gamma_R^{\text{ABS}}$. Negative differential conductance is observed when a quasiparticle is trapped in the weakly coupled BCS continuum such that transport is blocked when this pathway is energetically available. \textbf{(c)} Energy schematics for the hybrid quantum dot and normal leads. \textbf{Blue cross:} Electron transport is blocked at small source-drain bias due to presence of a superconducting gap in the hybrid quantum dot. \textbf{Red triangle:} Transport through an ABS level at small source-drain bias. \textbf{Orange circle:} At larger source-drain bias electron transport is blocked when a quasiparticle is trapped in the weakly coupled BCS continuum due to a slow relaxation time between the bound state and the continuum, leading to negative differential conductance. \textbf{Black inverted triangle:} Transport through both ABS levels and BCS continuum. The full transport process involves both pair splitting and recombination, and is detailed fully in Supplementary Section II. \textbf{Green square:} Transport through both ABS levels and BCS continuum.}
    \label{fig:3}
\end{figure*}

The device we consider consists of a \qty{100}{\nano\meter} diameter semiconductor InAs nanowire with a wurtzite crystal structure on which three different sections are defined along its length, see Fig.~\ref{fig:2}a. The main proximitized section (D1) has Al epitaxially matched on two of the six side facets positioned next to an array of permalloy micromagnets. Of the two control sections without micromagnets, one has been proximitized (D3) while the other is bare (D2). To ensure that the nanowire is within nanometre distance of the micromagnets, we used atomic force microscope (AFM) manipulation to place the nanowire in position (within $\sim \qty{10}{\nano\meter}$ distance) before contacting by Ti/Au leads, see also Supplementary Section IV. Tunnelling between the nanowire and the leads is controlled using gate voltages $V_{t1}$, $V_{t2}$  on Ti/Au gate electrodes, while charge occupancy is controlled using $V_{\text{g}}$. An external magnetic field with a direction parallel to the micromagnets (perpendicular to the nanowire axis) is used to polarize the micromagnet array such that it generates a field that is perpendicular to the nanowire in its aligned configuration, or a rotating magnetic field in its antialigned configuration, see Fig.~\ref{fig:2}b.

The dimensions of the micromagnets were chosen to consists of alternating wide (\qty{140}{\nano\meter}) and narrow (\qty{100}{\nano\meter}) micromagnets of uniform length (\qty{1}{\micro\meter}) and thickness (\qty{90}{\nano\meter}). Since narrow micromagnets have higher coercive fields than wider ones, this allows us to prepare the antiparallel configuration: all micromagnets are first polarised in one direction using an external magnetic field which is then swept in the other direction until it surpasses the coercive field of the wider micromagnets, while remaining below that of the narrow micromagnets.

We simulated the magnetic field using a finite difference method ~\cite{vansteenkiste2014} that accounts for the specific material parameters of the permalloy micromagnets and their dimensions, see Methods Section for details. As shown in Fig.~\ref{fig:2}b, the magnetic field profile rotates along the nanowire with a strength of order \qtyrange{100}{150}{\milli\tesla} directly adjacent to the micromagnets that decreases rapidly with distance.

The micromagnet simulations furthermore yield expected coercive fields on the order of \qty{30}{\milli\tesla} for the wide micromagnets and 50 mT for the narrow micromagnets. To experimentally verify these results, we tested the micromagnet arrays using  magneto-optic Kerr effect (MOKE) measurements and magnetic force microscopy (MFM), as shown in Figs.~\ref{fig:2}c and \ref{fig:2}d, respectively. The MOKE measurements show a magnetization plateau around \qty{35}{\milli\tesla} corresponding to the antiparallel configuration. The transitions between the plateaus are not instantaneous, reflecting the range of possible coercive fields, in agreement with the detailed micromagnet simulations, see Supplementary Section III. All low-temperature transport measurements on the device presented here were obtained at the \qty{6}{\milli\kelvin} base temperature of a dilution refrigerator using strongly filtered dc lines, see Methods Section for experimental setup details.

\section{Andreev bound state transport spectroscopy}

To measure the effect of the micromagnets, we performed transport spectroscopy on the proximitized InAs nanowire  (section D1). Figure \ref{fig:3}a shows differential conductance measurements of the device as a function of gate voltage and source-drain bias voltage in the absence of an external magnetic field and with the micromagnets in an antiparallel configuration. The data reveals the presence of Coulomb blockade diamonds with charging energy $E_C$ of $\sim\qty{1.5}{\milli\electronvolt}$ which, given an Al bulk superconducting gap energy $\Delta \lesssim \qty{180}{\micro\electronvolt}$, implies that single-electron charging takes place in the regime where $\Delta \ll E_C$~\cite{averin1992}. Different from conventional charge stability diagrams - that is, without superconductivity present - current is suppressed at small bias and is further characterised by single excitation lines with positive slopes parallel to the Coulomb diamonds.

We show here that these observations are fully explained by an Andreev bound state transport model as demonstrated by the simulated conductance plots of Figs.~3a and 3b. The energy diagram schematics of Fig.~3c illustrate the transport mechanisms for the various positions in the charge stability diagrams as indicated by the symbols. Details of the model are provided in Supplementary Section II. In the data and model, the observed subgap features depend on the strength and symmetry of the coupling between the Andreev bound states and the source and drain leads which, in our experiments, are tunable using the gate electrodes. For the data presented in Fig.~3a, and corresponding model calculations, this coupling is asymmetric. For the model calculations in the right plot of Fig.~3b - and further data presented in Fig.~\ref{fig:5} below - the coupling is symmetric. Importantly, in either situation, two key parameters can be extracted from the measurements: the proximitised superconducting gap energy $\Delta^*$ and the ABS energy $E_{\mathrm{A}}$ as shown by the labels in Fig.~\ref{fig:3}b. This in turn allows us to extract these parameters for the different micromagnet configurations and their evolution in an external magnetic field, as we describe below.

\begin{figure*}
    \centering
    \includegraphics[width=178mm]{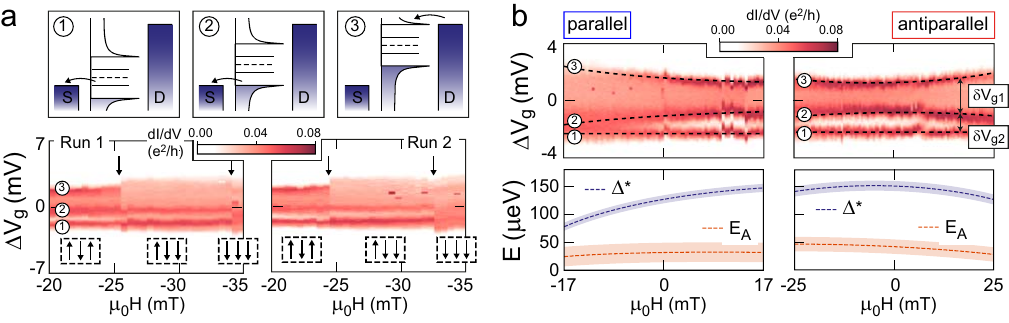}
    \caption{
    \textbf{Micromagnet array configurations and magnetic field dependence.}
    \textbf{(a)} The bottom panels show differential conductance as a function of applied gate voltage $\Delta V_{\text{g}}$ and external magnetic field for a fixed bias voltage $V_{\text{sd}}=\qty{0.44}{\milli\volt}$ for two different magnetic field sweeps. The three labelled transitions 1-3 correspond to those along the dashed black line in Fig.~3a. In both sweeps the transport data shows abrupt changes due to the switching of individual micromagnets within the array. The magnetisation direction of the three micromagnets along the nanowire is indicated by the boxed black arrows for the different magnetic field sweeps. The top panels show the energy schematics for the labelled transitions and the alignment of the bound states and superconducting gap edges with the electrochemical potentials of the source and drain electrodes.  \textbf{(b)} Differential conductance as a function of applied gate voltage $\Delta V_{\text{g}}$ and external magnetic field for a fixed bias voltage of $V_{\text{sd}}=\qty{0.44}{\milli\volt}$ for fixed parallel and antiparallel micromagnet configurations. The dashed black lines track the magnetic field dependence of the transition lines. The linegraphs in the bottom plots show the obtained superconducting gap energy $\Delta^*$ and Andreev bound state energy $E_A$ as a function of the external magnetic field. Graded areas indicate the estimated error bars.}
    \label{fig:4}
\end{figure*}

\section{Micromagnet array control}

To configure the micromagnet array in the desired configuration, we first polarise the micromagnets by applying a magnetic field of \qty{-200}{\milli\tesla} to align the magnetic moments in a parallel configuration. We then sweep the field up to \qty{+35}{\milli\tesla}, which induces a transition from parallel to antiparallel alignment. Subsequently, sweeping back from \qty{+35}{\milli\tesla} through \qty{0}{\milli\tesla} to \qty{-35}{\milli\tesla}, we observe the micromagnets switching direction from antiparallel back to parallel configuration. This is illustrated in Fig.~\ref{fig:4}a, which shows the differential conductance as a function of external magnetic field at a fixed bias of $V_{\text{sd}}=\qty{0.44} {\milli\volt}$ corresponding to the cross-section indicated by the black dashed line in the stability diagram of Fig.~\ref{fig:3}a. The magnetic field sweeps show individual micromagnets switching over two different, but nominally identical, runs. We find that the abrupt changes in the transport data, corresponding to the flipping of a single micromagnet, are reproducible with switching occurring within a range of \qtyrange{5}{15}{\milli\tesla}. This is consistent with our simulations presented in Supplementary Section III and the MOKE measurements presented in Fig.~\ref{fig:2}c. From the details of the transport data it is also possible to identify which micromagnet within the array is flipping. When a micromagnet flips to align with the external field, the resulting increased local magnetic field leads to a broadening of the superconducting gap edge in the density of states. In Fig.~\ref{fig:4}a, the transition labelled 3, which corresponds to transport between the continuum and the drain electrode, significantly broadens after the first micromagnet flip while the other transitions are unchanged. This broadening indicates that the micromagnet closest to the drain electrode has aligned with the field. Likewise, the transition labelled 2, which corresponds to transport between the continuum and the source electrode, broadens after the second micromagnet flip, suggesting that the micromagnet adjacent to the source is the second to switch its orientation. Full magnetic field sweeps are provided in Supplementary Section V.

\section{External magnetic field sweeps}

Having demonstrated control of individual micromagnets and array configurations, we determine how the superconducting gap and ABS energies change with an external magnetic field for both the parallel and antiparallel magnet configurations as shown in Fig.~\ref{fig:4}b. In all cases, the external field is directed parallel to the micromagnets' axes. The various energies are related as:
\begin{equation}
    2\Delta^* = e V_{\text{sd}}-\alpha e \delta V_{g1} \hspace{0.3cm} \textrm{and} \hspace{0.3cm} E_{\mathrm{A}}=\Delta^* - \alpha e \delta V_{g2}
\end{equation}
\noindent where in these measurement we again use a fixed bias of $V_{\text{sd}}=\qty{0.44} {\milli\volt}$ and where $\delta V_{g1}$ and $\delta V_{g2}$ are the differences in gate voltage between the transition lines as indicated in Fig.~\ref{fig:4}b. Since the gate lever arm $\alpha$ is also known - from the Coulomb diamond pattern we extract $\alpha=0.072$ eV/V - the energies $\Delta^*$ and $E_{\mathrm{A}}$ can be obtained. For the parallel micromagnet configuration, we then observe a magnetic field dependence of the proximitized superconducting gap energy $\Delta^*$ which varies continuously from $\sim 150$ $\mu$eV to \qty{75}{\micro\electronvolt} as the field is swept from positive to negative values, as shown in Fig.~\ref{fig:4}b. The antiparallel micromagnet configuration exhibits markedly different behaviour in an external field. For this configuration we observe, slightly offset from zero external field, a maximal superconducting gap energy $\Delta^*=\qty{150}{\micro\electronvolt}$ which decreases in either direction of the magnetic field sweeps.

Qualitatively, this can be understood from the alignment direction of the micromagnets with the external field. For the parallel configuration, the external field is aligned with the micromagnets for negative field directions: the external and micromagnet fields add, resulting in a small superconducting gap energy. For positive field directions, the external field and micromagnets are antialigned and the opposite effect occurs: the fields partially cancel which results in a larger superconducting gap energy. For the antiparallel micromagnet configuration - that is, a rotating field along the nanowire - every other micromagnet in the array is (mis)aligned with the external field for both sweep directions. Model calculations of transport spectroscopy in an external magnetic field are discussed in Supplementary Section V.

Similarly to the superconducting gap energy, we can extract the ABS energy $E_{\mathrm{A}}$ from the data, albeit with quite some uncertainty for this data set as this is based on the previous estimate from the data of $\Delta^*$. We obtain $E_{\mathrm{A}}=30 \pm 20$ $\mu$eV for parallel configuration and $45 \pm 20$ $\mu$eV for the antiparallel configuration which remains mostly unchanged over the (relatively small) external magnetic field range of the measurements in Fig.~4b.

\section{Even-odd parity filling}

\begin{figure*}
    \centering
    \includegraphics[width=172mm]{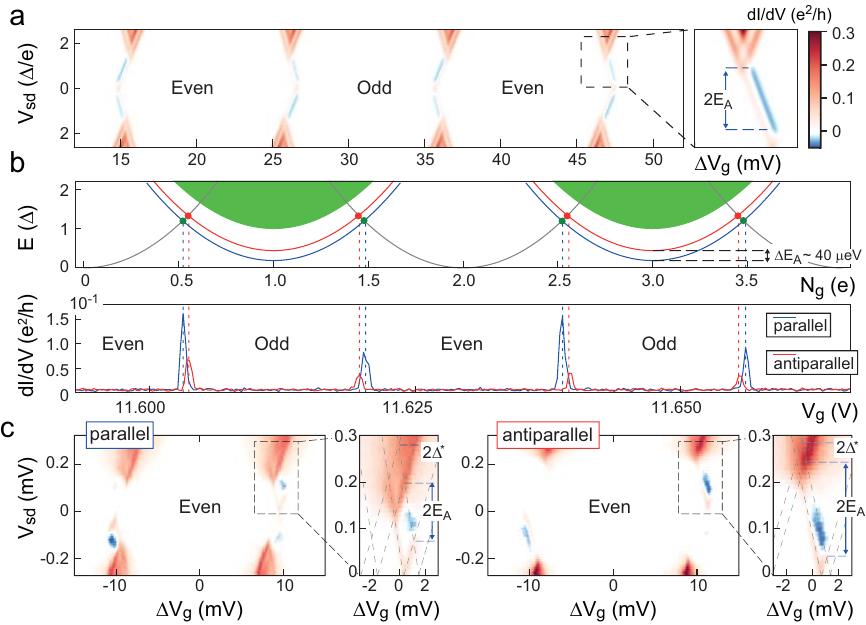}
    \caption{
    \textbf{Subgap transport spectroscopy and even-odd parity filling.}
    \textbf{(a)} Modelled differential conductance as a function of gate voltage $V_{\text{g}}$ and source-drain bias voltage $V_{\text{sd}}$ for a hybrid quantum dot symmetrically coupled to source and drain electrodes. The low-bias transport shows an alternating spacing of conductance features dependent on the parity of the dot occupation, as well as negative differential conductance. The position and length of the negative differential conductance regions are a direct measure of the ABS energy $E_A$ as shown in the inset. Details of the model parameters are provided in Supplementary Section II \textbf{(b)} Top: Energy dispersion as a function of gate charge. The different parabolas correspond to different charge occupancies $N$ of the hybrid quantum dot as indicated. For odd occupancies the parabolas are offset by the ABS energy $E_A$. The red and blue curves show this offset for two different ABS energies. The solid green curves correspond to the BCS continuum. Bottom: Measured conductance as function of gate voltage $V_{\text{g}}$ for the micromagnets prepared in a parallel (blue) and antiparallel (red) configuration, showing that the parallel configuration has an ABS energy lower by $\Delta E_A \sim 40$ $\mu$eV.
    \textbf{(c)}. Measured differential conductance as a function of gate voltage $V_{\text{g}}$ and source-drain bias voltage $V_{\text{sd}}$ for the parallel (left) and antiparallel (right) micromagnet configurations. Scale bars as in panel (a). The data shows clear even-odd parity filling in both cases and a distinct difference in the length of the regions of negative differential conductance from which the ABS energies are obtained.}
    \label{fig:5}
\end{figure*}

Next, we use the ability to control the coupling of the hybrid quantum dot to its leads via gates $V_{t1}$ and $V_{t2}$ to study the device in a symmetrically coupled transport regime. Figure \ref{fig:5}a shows the expected (modelled) low-bias transport in this regime which is characterised by an even-odd parity filling and negative differential conductance. Both transport features are observed in our measurements. The even-odd parity filling is apparent, for example, in the zero-bias measurements, shown for both the parallel and antiparallel magnet configurations in Fig.~\ref{fig:5}b. In both cases the even-odd alternation can be understood considering the free energy as a function of the induced gate charge $N_g = C_g V_g/e$, where $C_g$ is the gate capacitance, for different charge occupancies of the hybrid quantum dot, corresponding to the different parabolas. For odd charge occupancies, the parabolas are offset by the energy $E_A$ of the Andreev bound state. As zero-bias conductance peaks are observed when parabolas cross - that is, when the charge occupancy of the quantum dot is allowed to fluctuate - this leads to the observed even-odd pattern \cite{higginbotham2015}. Important in the context of our work is that since this offset is set by $E_A$, it is controlled by the micromagnet configuration as shown by the red and blue curves in Fig.~\ref{fig:5}b. From the spacings of the conductance peaks in the data the difference in Andreev bound state energies between the magnet configurations can therefore be obtained directly, yielding $\Delta E_A = 40 \pm 10$ $\mu$eV.

The Andreev bound state and superconducting gap energies can also be extracted from the differential conductance measurements shown in Fig.~\ref{fig:5}c for both the parallel and antiparallel magnet configurations. As detailed above, the negative differential conductance observed in both measurements can be understood as resulting from the sequential filling of an Andreev bound state: as a quasiparticle enters the continuum, it blocks electron transport through the device - for sufficiently long quasiparticle lifetimes, see Supplementary Section II - until it relaxes into the bound state, resulting in negative differential conductance \cite{higginbotham2015}. The bias window length of the negative differential conductance region is equal to $2E_A$ from which we obtain $E_{\mathrm{A}}=60 \pm 5$ $\mu$eV for the parallel configuration and $100 \pm 5$ $\mu$eV for the antiparallel configuration, in agreement with the difference in ABS energy obtained from the zero-bias measurements.

The larger ABS energy for the antiparallel configuration is consistent with expectations and with the measurements in the asymmetrically coupled transport regime discussed above: while our tight-binding simulations provide too simple a model to allow for a precise quantitative comparison, all simulations (see also Supplementary Section I) show a weaker dependence of the ABS energies on the micromagnet strength for rotating magnetic field textures as compared to the linear fields. Consistent with our results, the simulations show a difference of $\Delta E_A$ in the range of 10-40 $\mu$eV, depending on the precise value of the field of the micromagnets used ($\sim$ 100-150 mT) and $g$-factor ($\sim$ 5-10) of the InAs nanowire \cite{csonka2008}. From this observation, together with the external magnetic field dependence presented above, which shows a reduction of the superconducting gap energy in both field directions, we conclude that the antiparallel micromagnet configuration realises a rotating magnetic field texture along the proximitized InAs nanowire. From the lithographically defined pitch of $R=480$ nm for the micromagnet array and the InAs effective mass of 0.023 $m_e$, we estimate \cite{kjaergaard2012} the effective synthetic Rashba coefficient $\alpha_{\mathrm{eff}}=\hbar/2mR$ to be around $0.022$ eV nm, which is comparable to InAs's intrinsic spin-orbit coupling strength.

\section{Conclusions and Outlook}

Our work experimentally demonstrates the coupling of a proximitized InAs superconductor-semiconductor nanowire to micromagnet arrays. Using an external magnetic field, we are able to controllably configure the micromagnets to create a rotating magnetic field along the nanowire. This constitutes an experimental realisation of the engineered Hamiltonian of the kind introduced in Eq.~1 which yields a synthetic spin-orbit interaction term and manifests itself as a difference in the ABS energies for the different magnetic field textures. To additionally reach a topological non-trivial phase requires $g \mu_B B_0 \gtrsim \Delta$ \cite{oreg2010,kjaergaard2012}. While we achieve significant local magnetic fields of order 100-150 mT along the nanowire by placing them in very close proximity to the micromagnets using AFM nanomanipulation techniques, these fields are still about a factor of six or seven too low assuming $\Delta=180$ $\mu$eV and $g=10$, as also illustrated in Fig.~1b. Using magnetic materials such as cobalt instead of permalloy would be relatively straightforward but only improve field strengths by a factor of two. Stronger alternatives such as rare-earth micromagnets might therefore be needed and further developments to integrate such micromagnets with nanoscale devices would be of considerable interest. Alternatively, our work is readily applied to other semiconductor nanowires such as InSb and PbTe. The former is of interest because of the very high $g$-factors, of order 70, that have been observed ~\cite{nilsson2009,ophetveld2020} and which therefore allow for lower-strength magnets, while the latter has shown ballistic transport and conductance quantization, indicative of low disorder ~\cite{jung2022,zhang2023,wang2023,gupta2024}.

%%%%%%%%%%%%%%%%%%%%

\section{Methods}

\subsection{Device fabrication}
\qty{50}{\nano\meter}/\qty{3}{\nano\meter} of permalloy/\ce{Al2O3} micromagnet arrays were deposited using thermal evaporation on an electron-beam lithographically patterned doped \ce{Si}/\ce{SiO2} substrate with \qty{300}{\nano\meter} of oxide.
InAs nanowires of diameter \qty{100}{\nano\meter} were grown with a wurtzite crystal structure with Al epitaxially matched on two of the six side facets following the method described in Refs.~\cite{krogstrup2015,chang2015}. Using a micromanipulator, they were deposited near prepatterned micromagnets.
Proximitised superconducting sections as well as the non-superconducting sections of the nanowire were lithographically patterned followed by a chemical etch in Al Etchant D at \qty{20}{\celsius} for \qty{80}{\second}.
The etched nanowires were pushed closer to the micromagnet arrays using AFM nanomanipulation using NuNano Scout 350 probes of spring constant \qty{42}{\newton\per\meter}, limiting the tip velocities to \qty{10}{\nano\meter\per\second}. Al Etchant D can delaminate permalloy micromagnets, so it is important to perform the chemical etch before AFM nanomanipulation to reduce the risk to the prepatterned micromagnets.
After electron-beam lithographic patterning of the contacts and gates, native oxides are removed using Ar milling at \qty{2.5}{\pascal}, \qty{157}{\volt}, \qty{8}{\minute}, immediately followed by electron-beam evaporation of \qty{5}{\nano\meter}/\qty{95}{\nano\meter} of Ti/Au contacts and gates.

\subsection{Control measurements}
To verify that the switching events in magnetic field can be reliably attributed to the presence of micromagnets, control sections without magnets were fabricated on the same InAs nanowire. In particular, device control section D3 is of the same length and is proximitized in the same way as the main device section D1 but is not coupled to a micromagnet array. The control section shows similar charging energies and Andreev bound state transport as observed for the device section presented in the main text. This shows that these are robust and reproducible features of our hybrid nanowire device. At the same time, the magnetic field dependence of the control section does not show any switching events; see data in Supplementary Section V.

\subsection{Measurement setup}
Measurements were done in a Bluefors LD dilution refrigerator system with a base temperature of 6 mK. The DC lines to the device were filtered using Tusonix 4209-053LF pi-filters (low-pass configuration) with 20 dB attenuation at \qty{10}{\mega\hertz} at room temperature, copper powder filters~\cite{mueller2013} with \qty{50}{\deci\bel} of attenuation at \qty{10}{\giga\hertz} at the \qty{50}{\kelvin} stage, and at the mixing chamber stage Minicircuits LFCN RL filters with cutoff frequencies of \qty{5}{\giga\hertz}, \qty{1.45}{\giga\hertz}, and \qty{80}{\mega\hertz}, RC filters with cutoff frequencies of \qty{159}{\kilo\hertz} and \qty{48}{\kilo\hertz}, and finally RC filters on the device PCB with a cutoff frequency of \qty{34}{\kilo\hertz}. The effective cutoff frequency for all of these stages of filtration is \qty{6.6}{\kilo\hertz} yielding an electron temperature of $\sim$ 12 mK.
Two-terminal differential conductance measurements were made using an SR830 lock-in amplifier to measure the output signal after amplification through an SR570 current preamplifier. DC voltages were supplied using an iTest BILT BE2142.

\subsection{Transport model}
We simulate transport in our device using a master equation based on Section 5 of the Supplementary Information to \citeauthor{higginbotham2015}~\cite{higginbotham2015}.
We use a model that simulates transport in a superconducting device with two leads, a superconducting continuum, and an Andreev bound state, consistent with experimental data in both regimes presented in the main body of the text. It uses a master equation approach, adjusting the quasiparticle distributions in the superconducting continuum based on the free energy difference between odd and even states. The steady-state occupation probabilities are determined using the Pauli master equation, which includes tunnelling rates between the leads and the BCS continuum, between the leads and the Andreev bound state, and relaxation processes between the BCS continuum and the ABS. Using the model, we calculate the current through the island using these probabilities and rates and can explain the observed asymmetry in conductance due to suppressed conductance between the bound state and the drain electrode. More details of the model are provided in Supplementary Section II.\\

\subsection{Micromagnet simulations}
The micromagnet simulations in Figs.~\ref{fig:2}b were performed using the MuMax3 software package~\cite{vansteenkiste2014}. We used parameters of $M_{\mathrm{sat}}=\qty{800}{\kilo\ampere\per\meter}$ for the saturation magnetisation and $A=\qty{13}{\pico\joule\per\meter}$ for the exchange constant, which are typical for permalloy, and a damping constant $\alpha=0.1$ which ensures fast convergence with minimal impact on the results. The simulations were performed using a cell size of \qtyproduct{10x10x5}{\nano\meter} arranged in a \numproduct{80x384x24} grid.

The finite-difference software furthermore allows us to simulate the switching field for the micromagnets given our micromagnet geometries. The resulting magnetisation for an average of 100 simulations is shown in Supplementary Section III, which shows switching fields centred around \qty{30}{\milli\tesla}. The width and height of these micromagnets were normally distributed with a standard deviation of \qty{5}{\nano\meter} to mimic fabrication variability. Because the coercive field of individual micromagnets within the micromagnet array depends on differences in the dimensions of the individual micromagnets, the orientation of nearby micromagnets, as well as the internal magnetisation distributions of the micromagnets known as c-states and s-states~\cite{rave2000,liu2004}, there is a range of coercive fields, even for identical arrays of micromagnets.

\section{Data availability}

The data that support the findings of this study are available via Zenodo at https://doi.

\section{Code availability}

The code used to model the transport properties of a superconducting island with a single Andreev bound state within the superconducting gap as well as the code used to simulate the micromagnet field profiles are available via Zenodo at https://doi.

\section{Acknowledgements}

The authors thank Karsten Flensberg and Andrew Higginbotham for sharing code to perform transport simulations with the rate equation model. M.P.H. and M.R.B. acknowledge support by the Engineering and Physical Sciences Research Council under Grant EP/L015242/1. K.M. and M.R.C. acknowledge support by the QuantERA grant MAGMA. K.M. acknowledges the support by the German Research Foundation under grant 491798118, the financial support by the Ministry of Economic Affairs, Regional Development and Energy within Bavaria’s High-Tech Agenda Project ``Bausteine für das Quantencomputing auf Basis topologischer Materialien mit experimentellen und theoretischen Ansätzen'' (Grant No. 07 02/686 58/1/21 1/22 2/23) and by the German Federal Ministry of Education and Research (BMBF) via the Quantum Future project ‘MajoranaChips’ (Grant No. 13N15264) within the funding program Photonic Research Germany. J.N. acknowledges support by EU FETOpen grant no. 828948 and the Danish National Research Foundation (DR101).

\section{Author contributions}
InAs nanowires were grown by T.K. and placed on the Si/SiO$_2$ substrates by A.V. under supervision of J.N. Permalloy micromagnets were fabricated by M.P.H., D.B. and J.C.G. under supervision of W.R.B. All other device fabrication, including electron-beam lithography, metal evaporation of Ti/Au electrodes, Al etching, and AFM manipulation of the InAs nanowires were done by M.P.H. MOKE measurements were done by D.B. under supervision of M.R.C. and W.R.B. Low-temperature transport measurements were done by M.P.H. with support of B.J.V. and under supervision of M.R.B. MuMax3 and transport model simulations were done by M.P.H. and K.M. with input from M.R.B. and M.R.C. Kwant simulations were done by K.G. with input from M.P.H and K.M. The overall project was supervised by M.R.B. and M.R.C. The manuscript was written by M.P.H., M.R.B. and M.R.C. with comments from all authors.

\section{Competing interests}

The authors declare no competing interests.

%\bibliography{bibliography}

%

%%%%%%%%%%%%%%%%%%%

\clearpage

\setcounter{secnumdepth}{1}

%\onecolumngrid
\setcounter{equation}{0}
\setcounter{figure}{0}
\setcounter{table}{0}
\setcounter{page}{1}
\setcounter{section}{0}

\makeatletter
\renewcommand{\thesection}{S\arabic{section}}
\renewcommand{\theequation}{S\arabic{equation}}
\renewcommand{\thefigure}{S\arabic{figure}}
\renewcommand{\bibnumfmt}[1]{[S#1]}
\renewcommand{\citenumfont}[1]{S#1}

\renewcommand\onecolumngrid{% <<<<<<
\do@columngrid{one}{\@ne}%
\def\set@footnotewidth{\onecolumngrid}% <<<<<<<<<<<<<<<<
\def\footnoterule{\kern-6pt\hrule width 1.5in\kern6pt}%
}

\onecolumngrid

\section*{Supplementary Information - Synthetic spin-orbit coupling in superconductor-semiconductor hybrid nanowires with micromagnet arrays:}

\noindent The supplementary information comprises the following sections:\\
\\
S1: Tight-binding simulations \\
S2: Andreev bound state transport model simulations\\
S3: Magnetic field profile and hysteresis simulations\\
S4: Device images and nanowire alignment\\
S5: Magnetic field sweeps and control measurements

\section{Tight-binding simulations}

\begin{figure*}[b]
    \centering
    \includegraphics[width=\linewidth]{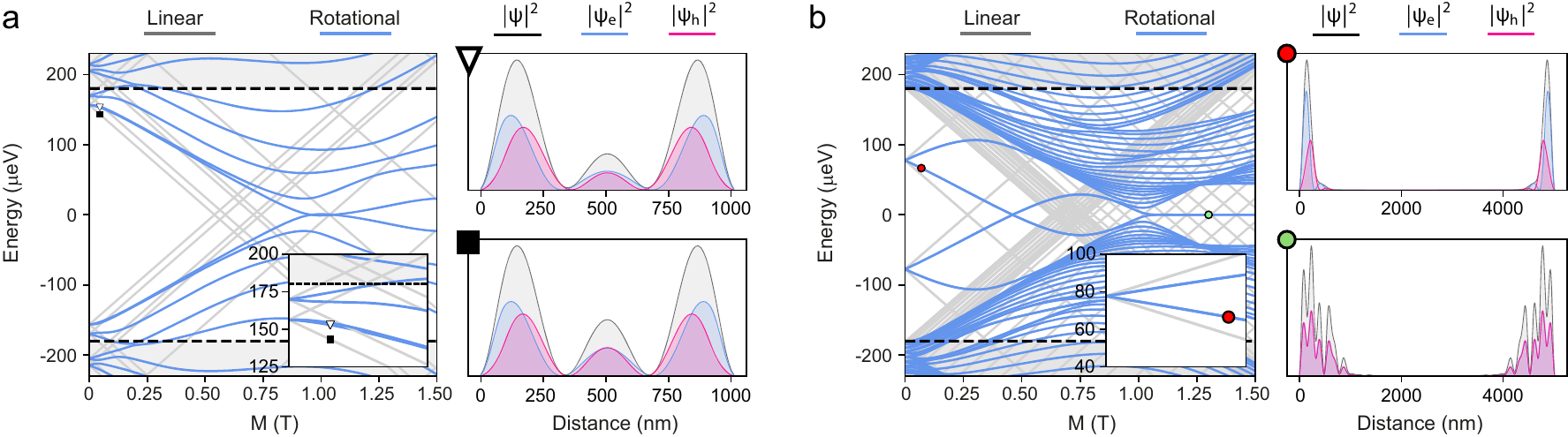}
    \caption{
    \textbf{Tight-binding simulations of hybrid superconductor-semiconductor nanowire coupled to a magnetic stray field texture.}
    \textbf{(a)} Excitation spectrum $E_{n}(\unit{\micro\electronvolt})$ as a function of the magnitude of the stray field texture $M(\unit{\tesla})$ for a \qty{1}{\micro\meter} nanowire. Inset: Sub-gap states for weak stray field. Probability densities $|\psi_{e}|^{2}$ (electron), $|\psi_{h}|^{2}$ (hole), and $|\psi|^{2}$ (total) of sub-gap states are shown for antiparallel (\textbf{triangle}) and parallel (\textbf{square}) configurations, illustrating the significant hybridisation between ABS for a wire with lengths $L_{\textrm{InAs}} = \qty{1}{\micro\meter}$ and $L_{\textrm{Al}} = \qty{0.8}{\micro\meter}$.
    \textbf{(b)} Excitation spectra $E_{n}(\unit{\micro\electronvolt})$ vs. stray field magnitude $M(\unit{\tesla})$ for a longer wire with $L_{\textrm{InAs}}= \qty{5}{\micro\meter}$ and $L_{\textrm{Al}} = \qty{4.6}{\micro\meter}$ for rotating (blue) and linear (grey) stray fields. Inset: Sub-gap states for weak stray field. Probability densities $|\psi_{e}|^{2}$ (electron), $|\psi_{h}|^{2}$ (hole), and $|\psi|^{2}$ (total) shown for eigenstates (colored dots): \textbf{Red}: Weak rotational ($M = \qty{80}{\milli\tesla}$), \textbf{Green}: Strong rotational ($M = \qty{1.3}{\tesla}$). Dashed lines represent the bulk superconducting gap of the alumnium $\Delta_{\textrm{Al}} = \qty{180}{\micro\electronvolt}$. Simulation parameters: tight-binding lattice constant $a = \qty{10}{\nano\meter}$, $L_{\textrm{Al}} = \qty{4.6}{\micro\meter}$, $m_{\mathrm{eff}} = 0.023m_{e}$, $\mu = \qty{0}{\micro\electronvolt}$, $\alpha_{\textrm{InAs}} = \qty{0.03}{\electronvolt\nano\meter}$, $\Delta^* = \qty{180}{\micro\electronvolt}$, $g_{\textrm{InAs}} = 10$ and $R = \qty{480}{\nano\meter}$.}
    \label{fig:S1}
\end{figure*}

We employ a one-dimensional tight-binding model using the Kwant [1] software package to simulate the excitation spectrum and wavefunctions of a superconductor-semiconductor nanowire in the presence of micromagnet-induced fields, as shown in Fig.~\ref{fig:S1}.
We simulated a device consisting of a \qty{5}{\micro\meter} long InAs semiconducting nanowire, with a \qty{4.6}{\micro\meter} long central portion proximitized by an aluminum shell, inducing superconductivity, and a shorter wire of similar dimensions to that of the experimental device of \qty{1}{\micro\meter} nanowire length and \qty{0.8}{\micro\meter} superconducting section. In both cases, gold contacts cap both ends of the wire. We compare the effects of two distinct magnetic field configurations: a linear field perpendicular to the wire axis and a rotating field, which we model using the rotating vector $\vec{M}(x) = M[\sin(2 \pi x / R) , \cos(2 \pi x/R) , 0 ]$ for a rotation period of $R = \qty{480}{\nano\metre}$.\\
\\
\noindent The model reveals the presence of a pair of ABSs, each originating in the normal regions between the superconductor and the metallic contacts, with each of these interfaces having a finite reflection constant. These bound states exhibit different behaviours under the two magnetic field configurations. In the linear field case, the bound states split with increasing field strength, following the expected Zeeman energy for a g-factor of 10. In contrast, the rotating field induces a significantly reduced splitting of the bound states. Moreover, as the rotating field strength increases, these states display a level-repulsion behaviour with the closing continuum states at the superconducting gap edge. In both short (\qty{1}{\micro\meter}) and long (\qty{5}{\micro\meter}) nanowires, the ABS wavefunctions initially localize in the NNS regions at the wire ends. However, due to wavefunction overlap in the middle, these states hybridize into bonding and antibonding configurations. The bonding/antibonding splitting is pronounced in the short nanowire, resulting in a single observable (bonding) ABS in the excitation spectrum. In contrast, this splitting is minimal in the long nanowire. Consequently, the ABS wavefunction extends across the entire length of the short nanowire, while remaining localized at the ends in the long nanowire. These states exhibit both electron and hole components of differing magnitudes, confirming their Andreev bound state nature in both linear and rotating micromagnetic configurations.\\
\\
\noindent Examining the case of a larger rotating magnetic field, of order \qty{1.3}{\tesla}, we observe the emergence of a zero-energy bound state. The wavefunction of this state is again localized at the ends of the nanowire, although with largest magnitude of the wavefunction located at the end of the superconductor, and with identical electron and hole components. This state corresponds to a Majorana zero mode. Compared to the Andreev bound states shown in Figure \ref{fig:S1}b (red dot), the Majorana zero mode has a larger localisation length, which is indicative of the lower energy gap in the topological phase as shown in the spectrum plot of \ref{fig:S1}b. Figure 1b in the main text is a simplified version of the same figure shown here in Fig.~\ref{fig:S1}.

\section{Andreev bound state transport model simulations}

We follow the rate equation model described in Section 5 of the Supplementary Information to Higginbotham \textit{et al} [2] to simulate transport in our device which models a superconducting island with a single Andreev bound state within the superconducting gap. Additional to Ref.~[2] we also take into account possible asymmetries in the coupling of the Andreev bound states to the source and drain leads.\\
%\subsection{Particle distributions}
\\
\noindent \textbf{Particle distributions}: The distribution of electrons in the lead is given by the normal Fermi-Dirac distribution centered about a chemical potential $\mu$:

\begin{equation}\label{eq:fermidirac}
    f_N(E) = \frac{1}{e^{\beta (E - \mu)}+1}
\end{equation}
\\
where $\beta=1/k_{\mathrm{B}} T$ denotes the inverse temperature of the lead. When considering the distribution of quasiparticles in the continuum of the superconductor we find that this must be modified, depending on the parity of the number of quasiparticles. The reason for this is that there is a difference in free energy $\delta F_{\mathrm{BCS}}$ between odd and even states, as discussed in Higginbotham \textit{et al} [2]. The free energy difference is given by [3]:

\begin{equation}
    \delta F_{\mathrm{BCS}} \approx \Delta - \beta^{-1} \ln{N_{\mathrm{eff}}}
\end{equation}
\\
where $N_{\mathrm{eff}}$ is the effective number of quasiparticle states. This free energy modifies the energy found in equation \ref{eq:fermidirac} giving the following expressions, where the second equality only holds at low temperature.
\begin{align}
    f_{\mathrm{e}} &= \frac{1}{e^{\beta (E + \delta F_{\mathrm{BCS}})}+1} \approx  N_{\mathrm{eff}} e^{-\beta (E + \Delta)} \\
    f_{\mathrm{o}} &= \frac{1}{e^{\beta (E - \delta F_{\mathrm{BCS}})}+1} \approx \frac{1}{N_{\mathrm{eff}}} e^{-\beta (E - \Delta)}
\end{align}
\\
\noindent The distribution of quasiparticles in the superconducting island hosting subgap states must consider the parity of the quasiparticle contributions of both the continuum and the subgap state. For $N$ electrons on the island, where the subgap state has occupancy $x \in {0, \uparrow, \downarrow, 2}$, the distribution of quasiparticles on the island is given by
\begin{equation}
    f_{S}(E,N,x) = \begin{cases}
        f_{\mathrm{e}}(E), \quad \text{for $N$ even and $x=0,2$},\\
        f_{\mathrm{o}}(E), \quad \text{for $N$ even and $x=\uparrow,\downarrow$},\\
        f_{\mathrm{o}}(E), \quad \text{for $N$ odd and $x=0,2$},\\
        f_{\mathrm{e}}(E), \quad \text{for $N$ odd and $x=\uparrow,\downarrow$}.
    \end{cases}
\end{equation}

\noindent \textbf{Steady state occupation probabilities}: The evolution of the occupation probability $P_{N,x}$ of a state parameterized by parameters $N$, the number of electrons on the island, and $x$, the occupancy of the subgap state is given by the Pauli master equation:
\begin{equation}\label{eq:probevolution}
    \frac{d P_{N,x}}{dt} = - \sum_{N',x'} \Gamma_{\substack{N'\leftarrow N \\ x'\leftarrow x}} P_{N,x} + \sum_{N',x'} \Gamma_{\substack{N\leftarrow N' \\ x\leftarrow x'}} P_{N',x'}
\end{equation}

\noindent where $\Gamma_{\substack{N'\leftarrow N \\ x'\leftarrow x}}$ is the rate between states $\ket{N,x}$ and $\ket{N',x'}$. In the steady state, the Pauli master equation for the probability $P_{N,x}$ is written
\begin{equation}\label{eq:steady_state}
    \frac{d P_{N,x}}{dt} = 0
\end{equation}

\noindent The rates in equation \ref{eq:probevolution} can be decomposed into contributions from tunneling between each lead ($\alpha \in \{L,R\}$) and the island, so they can be represented as

\begin{equation}
    \Gamma_{\substack{N'\leftarrow N \\ x'\leftarrow x}} = \Gamma_{\substack{N'\leftarrow N \\ x'\leftarrow x}}^L + \Gamma_{\substack{N'\leftarrow N \\ x'\leftarrow x}}^R
\end{equation}

\noindent and for the case where $x'=x$, we have

\begin{equation}
    \Gamma_{\substack{N'\leftarrow N \\ x'\leftarrow x}}^\alpha = \Gamma_{\substack{N'\leftarrow N \\ N_S-1 \leftarrow N_S}}^{\alpha,x} + \Gamma_{\substack{N'\leftarrow N \\ N_S+1 \leftarrow N_S}}^{\alpha,x}
\end{equation}

\noindent where $\Gamma_{\substack{N'\leftarrow N \\ N_S' \leftarrow N_S}}^{\alpha,x}$ represents the tunneling rate between lead $\alpha \in \{ L,R \}$ and the continuum of the superconducting island, with the Andreev bound state occupation $x$ remaining constant. The number of excitations in the continuum $N_S$ has the same parity as the product of the parities of $N$ and $x$. This master equation has contributions from three types of tunneling rates: tunneling between the lead and the continuum of the superconducting island, tunneling between the lead and the Andreev bound state, and relaxation between the continuum and the bound state. According to Fermi's golden rule, the tunneling rate between the lead and the continuum of the superconductor is

\begin{equation}
    \Gamma_{\substack{N+\chi\leftarrow N \\ N_S-\chi' \leftarrow N_S}}^{\alpha,x} = \gamma_\alpha \int^\infty_\Delta \frac{E dE}{\sqrt{E^2 - \Delta^2}} f_N(E^C_{N+\chi} - E^C_{N+\chi} + \chi' E - \chi \mu_\alpha)f_{S,-\chi'}(E,N,x)
\end{equation}

\begin{figure*}[p]
\vspace{3cm}
    \centering
    \includegraphics[width=158mm]{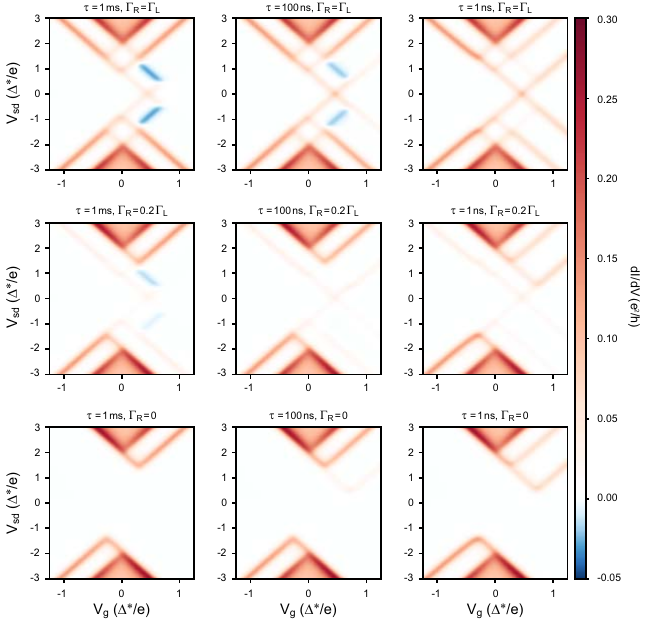}
    \caption{
    \textbf{Charge stability diagrams for various quasiparticle lifetimes and source-drain tunnel coupling asymmetries.}
    Differential conductance $dI/dV$ as a function of gate voltage $V_g$ and source-drain bias voltage $V_{sd}$ for nine different cases, parametrised by the quasiparticle lifetime in the subgap state $\tau=1/\Gamma_{relax}$ and the asymmetry $\Gamma_R/\Gamma_L$ of the tunneling rates to the leads as indicated. Further parameters are, $T=\qty{75}{\milli\kelvin}$, $\gamma_{\alpha}=0.09/\pi\hbar$ and $\Gamma_{L}=\qty{200}{\mega\hertz}$.}
    \label{fig:S2}
\end{figure*}

\noindent where $f_{S,+1}(E,N,x) = f_{S}(E,N,x)$ and $f_{S,-1}(E,N,x) = 1-f_{S}(E,N,x)$. $E^C_N$ refer to the charging energies of states $N$ in the constant interaction model. $\chi,\chi' \in \{ -1, 1 \}$ represent addition of an electron on the island and the addition of a quasiparticle to the continuum, respectively. $\gamma_\alpha$ is the coupling between the leads and the continuum, and is related to the normal state conductance by $g_{\text{Al}}=(\pi/2)(e^2/h)h\gamma_\alpha$. There is an additional factor of $h$ here compared to Ref.~[2] to ensure correct units.\\
\\
The tunneling rates between the lead and the subgap state are given by:

\begin{align}
    \Gamma_{\substack{N-1\leftarrow N \\ s\leftarrow 0}}^{\alpha}
    &\approx \Gamma_{\alpha}v_{0}^2 [1-f_{\mathrm{N}}(E^{c}_{N}-E_{s}-E^{c}_{N-1}-\mu_{\alpha})],\\
    \Gamma_{\substack{N+1\leftarrow N \\ s\leftarrow 0}}^{\alpha}
    &\approx \Gamma_{\alpha} u_{0}^2 f_{\mathrm{N}}(E^{c}_{N+1}+E_{s}-E^{c}_{N}-\mu_{\alpha}).\\
    \Gamma_{\substack{N-1\leftarrow N \\ 0\leftarrow s}}^{\alpha}
    &\approx \Gamma_{\alpha}u_{0}^2[1-f_{\mathrm{N}}(E^{c}_{N}+E_{s}-E^{c}_{N-1}-\mu_{\alpha})],\\
    \Gamma_{\substack{N-1\leftarrow N \\ 2\leftarrow s}}^{\alpha}
    &\approx \Gamma_{\alpha}v_{0}^2 [1-f_{\mathrm{N}}(E^{c}_{N}-E_{\bar{s}}-E^{c}_{N-1}-\mu_{\alpha})],\\
    \Gamma_{\substack{N+1\leftarrow N \\ 2\leftarrow s}}^{\alpha}
    &\approx \Gamma_{\alpha} u_{0}^2 f_{\mathrm{N}}(E^{c}_{N+1}+E_{\bar{s}}-E^{c}_{N}-\mu_{\alpha}),\\
    \Gamma_{\substack{N+1\leftarrow N \\ 0\leftarrow s}}^{\alpha}
    &\approx \Gamma_{\alpha}v_{0}^2 f_{\mathrm{N}}(E^{c}_{N+1}-E_{s}-E^{c}_{N}-\mu_{\alpha}),\\
    \Gamma_{\substack{N-1\leftarrow N \\ s\leftarrow 2}}^{\alpha}
    &\approx \Gamma_{\alpha}u_{0}^2[1-f_{\mathrm{N}}(E^{c}_{N}+E_{\bar{s}}-E^{c}_{N-1}-\mu_{\alpha})],\\
    \Gamma_{\substack{N+1\leftarrow N \\ s\leftarrow 2}}^{\alpha}
    &\approx \Gamma_{\alpha}v_{0}^2 f_{\mathrm{N}}(E^{c}_{N+1}-E_{\bar{s}}-E^{c}_{N}-\mu_{\alpha}).
\end{align}

\noindent where $u_{0}, v_{0}$ are BCS coherence factors which we set to $1/\sqrt{2}$ throughout our analysis, and $s \in \{\uparrow, \downarrow\}$, while $\bar{s}$ denotes the opposite of $s$. The final contributions to the master equation are from the relaxation processes:

\begin{equation}
    \Gamma_{\substack{N_o\leftarrow N_o \\ s\leftarrow 0}} = \Gamma_{\substack{N_e\leftarrow N_e \\ 2\leftarrow s}} = \Gamma_{\text{relax}}
\end{equation}

\noindent and their corresponding excitations rates\footnote{Relaxation/excitation rates with an odd number of electrons $N_o$ are swapped compared to the version in Higginbotham \textit{et al} [2] to ensure relaxation is energetically favourable.}.

\begin{equation}
    \Gamma_{\substack{N_o\leftarrow N_o \\ 0\leftarrow s}} = \Gamma_{\substack{N_e\leftarrow N_e \\ s\leftarrow 2}} = \Gamma_{\text{relax}} e^{-\beta (\Delta - E_{\mathrm{A}})}
\end{equation}

\noindent When we solve equation \ref{eq:steady_state} in the steady state, we obtain a set of probabilities $P_{N,x}$. The current through the island is then given by the expression:

\begin{equation}
    I=(-e)\sum_{N,xx'}\left(
\Gamma^{L}_{\substack{N+1\leftarrow N \\ x'\leftarrow x}}
-\Gamma^{L}_{\substack{N-1\leftarrow N \\ x'\leftarrow x}}
\right)P_{N,x}.
\end{equation}

\noindent The differential conductance $dI/dV$ is then obtained by numerically evaluating the difference in current by incrementally varying the source-drain bias voltage $V_{SD}$. Figure~\ref{fig:S2} shows differential conductance charge stability diagrams as a function of gate voltage and source-drain bias voltage for a temperature of $T=\qty{75}{\milli\kelvin}$. In these plots, we vary the quasiparticle lifetime in the subgap state $\tau=1/\Gamma_{relax}$ (left to right) and the asymmetry of the tunnelling rates to the leads, given by the ratio of $\Gamma_R/\Gamma_L$ (top to bottom). These two parameters change the low-bias conductance of the system, with the decrease in the quasiparticle lifetime leading to a reduction in the negative differential conductance and a decrease in the tunnelling rate $\Gamma_R$ suppressing low-bias transport.\\
\\
\noindent The potential drop at the tunnel barriers between the island and leads is set by their respective capacitive couplings. For symmetrical capacitive coupling of leads to the island we define the lead chemical potentials as $\mu_L=\frac{1}{2}eV_{sd}$ and $\mu_R=-\frac{1}{2}eV_{sd}$. For asymmetric capacitive coupling these ratios change and skewing of the Coulomb diamonds is observed. This is taken into account in the model calculations in Figs.~3a and 5a in the main text which directly compare to the data.\\

\begin{figure}
    \centering
    \includegraphics[width=166mm]{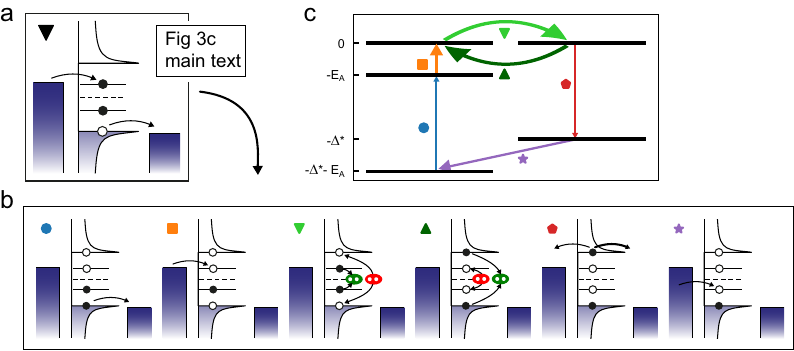}
    \caption{
    \textbf{Detailed breakdown of the transport in the case represented by the black inverted triangle of Fig.~\ref{fig:3}c in the main text.}
    \textbf{(a)} Simplified transport diagram as shown in the main text. \textbf{(b)} Detailed breakdown of the different steps involved in the transport. \textbf{Blue:} An electron-like quasiparticle tunnels out of the BCS continuum into the right lead. \textbf{Orange:} An electron tunnels from the left lead into the Andreev state. \textbf{Lime green:} A pair of electron-like Andreev quasiparticles form a Cooper pair (green), while a Cooper pair is split (red) forming a pair of electron-like quasiparticles in the BCS continuum. \textbf{Dark green:} The opposite process as seen in the lime green case. \textbf{Red:} An electron-like quasiparticle in the BCS continuum has transport allowed to both leads, with transport favoured in the forward direction. \textbf{Purple:} An electron tunnels into the Andreev state, returning the superconducting island to the initial state. \textbf{(c)} Energy level  diagram of the 5 different intermediate states involved in transport. The symbols and arrow colours correspond to transitions represented in \textbf{(b)}, and the size of the arrows correspond to the rates of different transport processes. The green arrow rates were not simulated, but Cooper pair splitting/recombination processes can be assumed to be faster than tunneling. Levels are separated to the left and right for clarity.}
    \label{fig:S3}
\end{figure}

\noindent \textbf{Parity effect}: Superconducting islands which are sufficiently small and isolated from their environment have their energy scales dominated by the Coulomb interaction and superconductivity. To formulate this description, we can write a Hamiltonian $H = H_{\text{C}} + H_{\text{BCS}}$, with:

\begin{equation}
    H_{\text{C}} = E_{\text{C}} \sum_N \left(N - N_{\text{g}}\right)^2
\end{equation}

\begin{equation}\label{eq:BCS_energy}
    H_{\text{BCS}} = \begin{cases}
        0 &\text{$N$ even}\\
        \Delta^* &\text{$N$ odd}
    \end{cases}
\end{equation}

\noindent In the case where an Andreev bound state exists, we can replace $\Delta^*$ in Equation \ref{eq:BCS_energy} with $E_{\mathrm{A}}$ because the ABS provides a lower energy level to fill than the superconducting continuum. The resulting parabolas given by this Hamiltonian are shown in Fig.~\ref{fig:5} in the main text, where we show parabolas corresponding to the superconductor and the Andreev bound states of both the parallel and antiparallel micromagnet configurations.\\
\\
At the points where these parabolas intersect, an island occupation of $N$ and $N+1$ are equally likely, and electron transport is allowed. This results in Coulomb blockade peaks with parity-dependent spacing, as is observed in the zero-bias conductance shown in Fig.~5 in the main text, for both parallel and antiparallel configurations, where we see different spacings. These data are obtained in the case of symmetric coupling to each of the leads, where non-negligible conductance is observed.\\
\\
\noindent \textbf{Detailed description of transport from Fig.~\ref{fig:3} in the main text}: To provide a more comprehensive understanding of the transport mechanisms, Fig.~\ref{fig:S3} presents a detailed breakdown of the case represented by the inverted black triangle in Fig.~\ref{fig:3} in the main text. Panel (c) presents an energy-level diagram depicting the five intermediate states involved in transport. The model did not explicitly simulate Cooper pair breaking and recombination processes.
However, this simplification is justified because these processes are energy-neutral with an appropriate pair of occupied or unoccupied  quasiparticle states with energies $\pm E_{\mathrm{A}}$ or $\pm\Delta$. We can reasonably assume that Cooper pair splitting/recombination processes occur at rates much faster than other relevant processes in the system, allowing us to treat them as instantaneous events in our simulations.

\section{Magnetic field profile and hysteresis simulations}

\begin{figure}
    \centering
    \includegraphics[width=172mm]{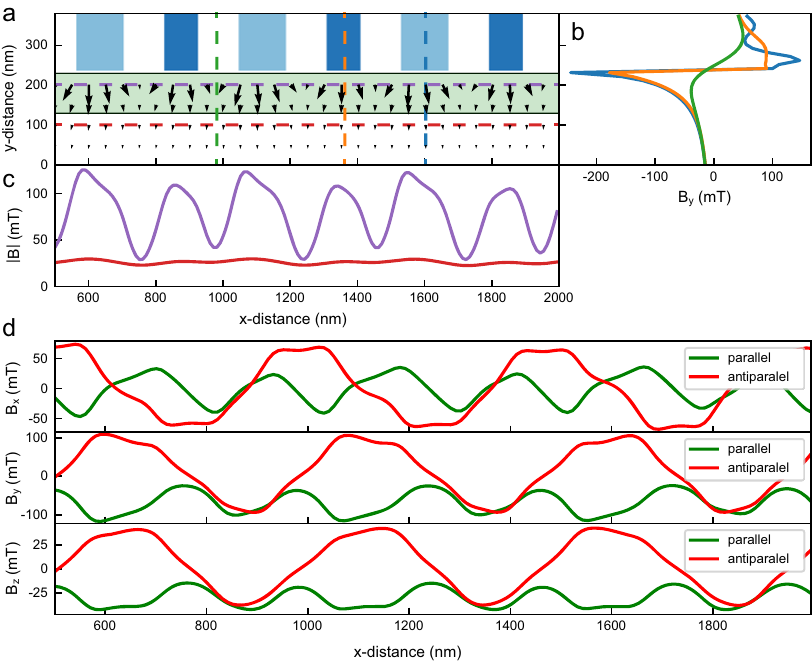}
    \caption{
    \textbf{Simulated magnetic field profile for parallel micromagnet configuration.}
    \textbf{(a)} Magnetic field profile of the same micromagnet array as shown in Fig.~\ref{fig:1} in the main text but in the parallel configuration, as simulated in MuMax3.
    One-dimensional cuts perpendicular to the wire \textbf{(b)}, and parallel to the wire \textbf{(c)} demonstrate that an oscillatory magnetic field of significant magnitude is only observable close to the micromagnet array. \textbf{(d)} Comparison of different components of the magnetic field generated by parallel (green) and antiparallel (red) micromagnet configurations. The peaks and valleys of the resulting oscillations, most visibly in the antiparallel configuration, are characterised by having pairs of lobes.
    }
    \label{fig:S4}
\end{figure}

The magnetic field profile for the parallel micromagnet configuration is shown in Fig.~\ref{fig:S4}a - similarly to that for the antiparallel configuration of Fig. 2a of the main text. Line traces of field along the dashed lines are shown in Fig.~\ref{fig:S4}b and c as indicated. For all magnetic field line traces the distance from the substrate (z-axis) is 55 nm, roughly through the center of the InAs nanowire.\\
\\
Figure \ref{fig:S4}d shows magnetic field line traces for both parallel (green) and antiparallel (red) micromagnet configurations. For all spatial directions the antiparallel configuration shows oscillatory behaviour around zero. The peaks and valleys of the resulting oscillations are characterised by pairs of lobes. The difference in height between each pair of lobes is evidence for spontaneous s-states and c-states in these simulations [4,5]. While the oscillatory behaviour is therefore not perfectly sinusoidal - as used in the simulations of Fig.~1 in the main text - the dominant frequency component of the field will be that corresponding to the pitch of the micromagnet array and it has been shown that the appearance of Majorana zero modes is robust against variations of the magnetic field profile [6,7]. The parallel micromagnet configuration also shows some oscillatory behaviour, but significantly offset from zero field for the $y$ and $z$-directions as expected. Indeed, due to the oscillatory component of the parallel magnet configuration (double that of the antiparallel one) it is also possible, in principle, to observe Majorana zero modes and a topological phase transition in this situation, albeit at rather different fields and with a Rashba spin-orbit term that changes sign along the wire [6].\\

\begin{figure}
    \centering
    \includegraphics[width=158mm]{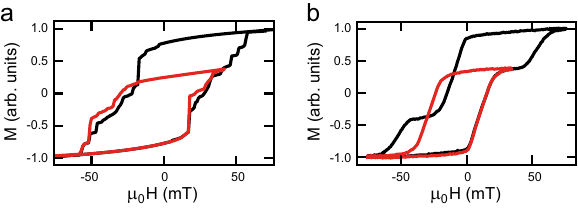}
    \caption{
    \textbf{Simulated micromagnet array hysteresis loop.}
    \textbf{(a)} Average of 100 simulated major (black) and minor (red) hysteresis loops of micromagnet arrays that include some randomness in their dimensions, modelling fabrication variability. The minor loop represents switching between the parallel and antiparallel micromagnet array configurations, while the major loop represents switching between two opposite parallel configurations, via the antiparallel one.\textbf{(b)} MOKE data reproduced from Fig.~\ref{fig:2}c in the main text. Coercive fields are consistent with the simulations presented in panel (a).
    }
    \label{fig:S5}
\end{figure}

\noindent \textbf{Micromagnet array hysteresis}: An array of alternating micromagnets of average width \qty{100}{\nano\meter} and \qty{140}{\nano\metre},  average length \qty{800}{\nano\meter}, and average thickness \qty{90}{\nano\meter} were simulated as shown in Fig.~\ref{fig:S5}a. A normally distributed error of standard deviation \qty{5}{\nano\meter} was added to each dimension of these micromagnets to model the effect of fabrication variability. To simulate a MOKE measurement, we simulate the effect of ramping an external magnetic field in a hysteresis loop from \qtyrange{-100}{100}{\milli\tesla}, and repeating the simulation 100 times. Repetition is important, as the actual MOKE measurements are also averaged over many copies of these micromanget arrays, and the effect of the randomized dimensions must be averaged out.\\
\\
\noindent Although the simulated MOKE measurement is not smooth, we observe micromagnet switching starting around \qty{20}{\milli\tesla}, and a small plateau appearing before the entire micromagnet array is polarized around \qty{60}{\milli\tesla}. This is consistent with the MOKE data shown in Fig.~\ref{fig:2}c of the main text. While some randomness was added artificially to mimic randomness introduced through imperfect fabrication, another source of randomness that occurs is due to the starting state of the fabrication. Through the hysteresis loop, the simulated micromagnets form s-states and c-states~[4,5], which randomize the instantaneous coercive fields for the individual micromagnets. This causes randomness over multiple repeats of even identical micromagnet arrays.

\section{Device images and nanowire alignment}

\begin{figure*}[p]
\vspace{2cm}
    \centering
    \includegraphics[width=178mm]{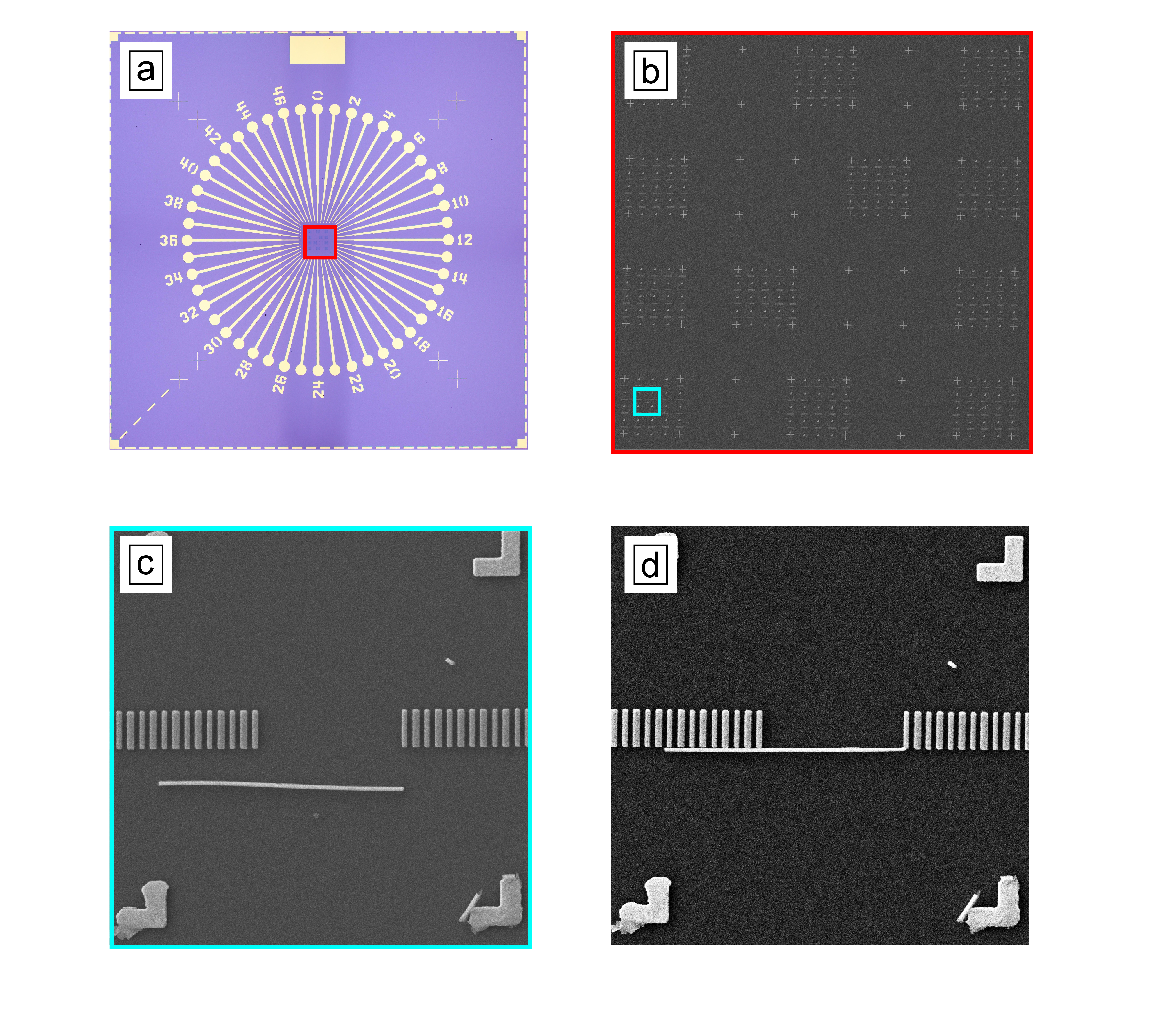}
    \caption{
    \textbf{(a)} Optical image of 4.5 mm x 4.5 mm size showing large scale features of a partially-fabricated chip. The numbered circles are the Ti/Au bondpads. There are eight cross-shaped alignment marks, two in each corner, which are essential for further lithography steps. \textbf{(b)} SEM image of the central region of the chip, with markings to help locate the deposited nanowires. Arrays of permalloy micromagnets were also deposited before the nanowires and visible in the image. \textbf{(c)} SEM image of the InAs/Al nanowire of Fig.~2 in the main text, with nearby micromagnet arrays. The nanowire was deposited by micromanipulator and therefore not precisely positioned at this stage. \textbf{(d)} SEM image of the InAs/Al nanowire following localised etching of the Al layer and the AFM nanomanipulation of the nanowire for precise positioning next to the micromagnets.}
    \label{fig:S6}
\end{figure*}

Supplementary Fig.~\ref{fig:S6} shows images of the device structure in the main text at various scales and stages of the fabrication process. Figure~\ref{fig:S6}a shows an optical image of the chip (boron-doped Si/SiO$_2$ substrate, of resistivity $< 0.004$ $\Omega$cm with 300 nm of oxide) on which the device is fabricated and with Ti/Au alignment markers and numbered bond pads deposited to connect the device to the sample holder. In the centre of the optical image, as indicated by the red square, sets of permalloy micromagnet arrays are subsequently fabricated, shown in detail by the SEM image in Fig.~\ref{fig:S6}b. The permalloy was capped with a 3 nm layer of Al$_2$O$_3$ to prevent oxidation of the magnets.\\
\\
\noindent InAs nanowires with Al epitaxially matched on two of the six side facets were then deposited on the substrate using a micromanipulator as shown in Fig.~\ref{fig:S6}c. At this stage the nanowires were lithographically patterned using a PMMA resist layer, followed by a chemical etch in Al Etchant D to remove the Al where desired, without the risk of delaminating the permalloy micromagnets. Before metal contacting, the nanowire is first moved to be in close proximity of the micromagnets using AFM nanomanipulations, as shown in Fig.~\ref{fig:S6}d. The completed device, including source, drain and gate electrodes is shown in Fig.~2a in the main text.

\section{Magnetic field sweeps and control measurements}

\begin{figure}
    \centering
    \includegraphics[width=170mm]{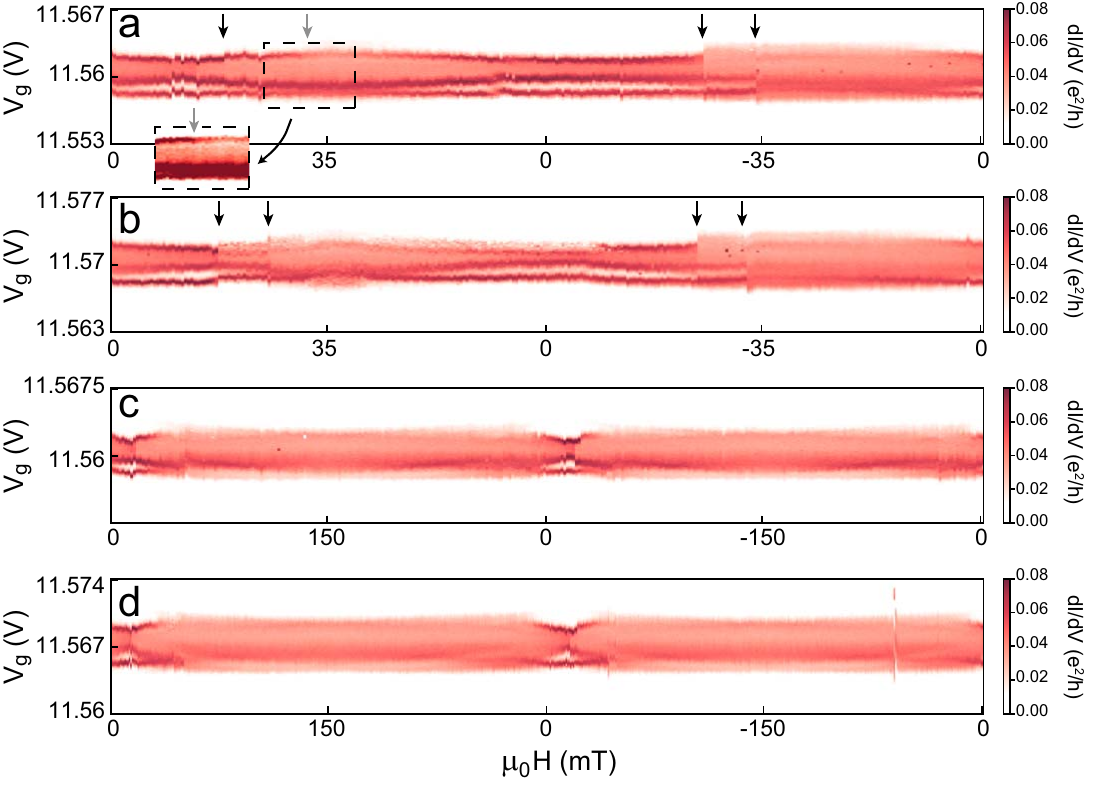}
    \caption{
    \textbf{Full magnetic field sweeps at high bias.}
    \textbf{(a,b)} Two repeated magnetic field sweeps between -35 and 35 mT. The micromagnet array was prepared in an aligned configuration prior to the sweep. A pair of micromagnet switching events are observed each time from the antiparallel to parallel transition. In the parallel to antiparallel transition, one switching event in the upper panel is not clearly observed. We tentatively identify this event to be around 32 mT shown by the grey arrow. The inset shows an enhanced contrast plot (full scale to 0.045 $e^2/h$) of the differential conductance for the area indicated by the dashed box, showing a change of conductance for the top transition at the magnetic field indicated by the arrow.
    \textbf{(c,d)} Two repeated magnetic field sweeps between -150 and 150 mT. Here we observe saturation of the signal at larger magnetic fields, due to the diminishing of superconductivity. The resulting quantum dot is less sensitive to micromagnet switching than the superconducting island that exists at low external magnetic fields.
    }
    \label{fig:S7}
\end{figure}

\begin{figure}
 \centering
     \includegraphics[width=90mm]{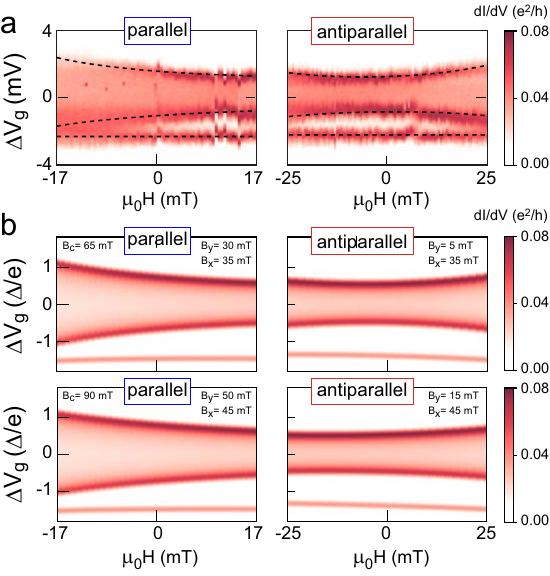}
     \caption{
     \textbf{External magnetic field model calculations.}
     \textbf{(a)} Differential conductance as a function of applied gate voltage $\Delta V_{\text{g}}$ and external global magnetic field for a bias voltage of $V_{\text{sd}}=\qty{0.44}{\milli\volt}$ and for fixed parallel and antiparallel micromagnet configurations. The dashed black lines track the magnetic field dependence of the transition lines. As shown in Fig.~4b of the main text. \textbf{(b)} Model calculations of the differential conductance as a function of applied gate voltage, in units of $\Delta/e$ and external global magnetic field for fixed parallel and antiparallel micromagnet configurations. Plots are shown for two sets of model parameters as indicated.}
     \label{fig:S8}
 \end{figure}

\noindent \textbf{Full magnetic field sweep data}: Fig.~\ref{fig:4}a in the main text shows cropped sections of two different magnetic field sweeps, shown in their entirety in Figs.~\ref{fig:S7}a-b here. These data were obtained using a peak tracking algorithm which adjusted for gate voltage $V_{\text{g}}$ drift, and additional post-processing to account for remaining voltage fluctuations was implemented. Thus, the gate voltage shown refers to the initial gate voltage used. Two further magnetic field sweeps up to a maximum magnetic field amplitude of \qty{150}{\milli\tesla} were done, which show saturation, indicating a complete closure of the superconducting gap. These are shown in Figs.~\ref{fig:S7}c-d.\\
\\
\noindent \textbf{Magnetic field sweep model}: The superconducting gap parameter $\Delta^*$ and the Andreev bound state energy $E_A$ are obtained directly from the data as discussed in the main text (Eq.~3). Here we discuss a simple model that - while not accounting for the full complexity of the magnetic field profile of the micromagnet array as shown above - provides a qualitative understanding of the overall dependence of the transport data in an external magnetic field. For our model we use the relations [2]:
\begin{equation}
   \Delta^* (B) = \Delta \sqrt{1-(B/B_c)^2} \qquad \textrm{and} \qquad  E_A^\pm(B)=\frac{\Delta^*(B)}{\Delta} E_A^0 \pm \frac{1}{2} g \mu_B B
\end{equation}

\noindent where $\Delta$ and $E_A^0$ are the superconducting gap and Andreev bound state energy in the absence of any magnetic field, respectively, and where $B_c$ is the critical field in perpendicular field. Here we assume $\Delta=180$ $\mu$eV and $g=10$. For the total field amplitude we use:

\begin{equation}
B=\sqrt{B^2_{x,m}+(B_{y,m}+B_{y,ext})^2+B^2_{z,m}}
\end{equation}
\noindent where the subscripts denote micromagnet and external fields in the $x$, $y$ and $z$ directions.\\
\\
\noindent To model the measured $\Delta^*(B)$ and $E_A(B)$ as a function of the external global field, we use the micromagnet field and the critical field $B_c$ of the Al superconductor as fit parameters. As shown in Fig.~\ref{fig:S8} this allows us to obtain good agreement between the model and experimental data for both parallel and antiparallel micromagnet configurations. While the extracted values should be considered with considerable care given the number of fit parameters ($B_c$, $B_{y,m}$ and $B_{x,m}$ - or $B_{z,m}$ as the model does not discriminate between these two directions) and model assumptions used, there are two robust features of the model. Firstly, the micromagnet field is sufficiently close to the critical field that the addition of an external global field has a significant effect on $\Delta^*$, which would not have been the case if $B/B_c \ll 1$. Secondly, in the absence of an external global field, the $B_{y,m}$ field  of the micromagnets (as seen by the superconducting Al half-shell) is smaller in the antiparallel magnet configuration as compared to the parallel configuration. As illustrated by the two examples shown in Fig.~\ref{fig:S8}b, this still leaves a range of fit parameters that provide reasonable agreement with the data. We find $B_c$ to be in the range of 60-120 mT with values for $B_{y,m}$ smaller than $B_c$ by about 40 mT for the parallel configuration and by about 60 mT for the antiparallel configuration. At the same time, $B_{x,m}$ and $B_{z,m}$ components are non-negligible, in the range of 30-40 mT. These field amplitudes are within the expected range of the micromagnet field simulations shown above - where we note that the position of the superconductor is extended over two facets of the InAs nanowire - and that the values for $B_c$ are consistent with previous findings [2].
\\

\begin{figure}
    \centering
    \includegraphics[width=138mm]{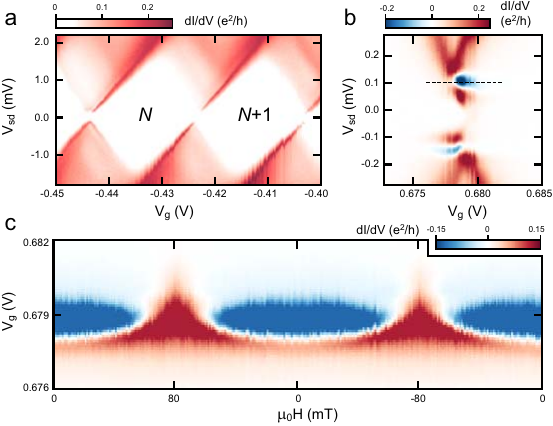}
    \caption{
    \textbf{Transport spectroscopy of control device.}
    \textbf{(a)} Coulomb diamonds measured in section D3, which has no micromagnets. \textbf{(b)} Low-bias transport spectroscopy displaying, indicative of the presence of Andreev bound states and long quasi-particle lifetimes. \textbf{(c)} Magnetic field dependence of the cut through the negative differential conductance shown in panel (b). No discontinuities are observed here without nearby micromagnets.}
    \label{fig:S9}
\end{figure}

\noindent \textbf{Control device measurements}: Two control devices (sections D2 and D3) were fabricated on the same hybrid nanowire, both of which had no adjacent micromagnets, as seen in Fig.~\ref{fig:2}a in the main text. Section D2 additionally had the Al half shell removed. Section D3 still had the Al half shell, and measurements on this section are shown in Fig.~\ref{fig:S9}. As shown in Fig.~\ref{fig:S9}a, a Coulomb blockade pattern is observed with a charging energy of approximately 1.7 meV. This is similar to that of section D1 (which is of the same length) and demonstrates that the hybrid quantum dot is defined by the device structure rather than, for example, a spurious defect-induced quantum dot. As expected, given the presence of the Al half shell in section D3, Andreev bound states and negative differential conductance are observed in transport measurements, albeit less well resolved at the gate voltages explored here as compared to section D1. Importantly, as shown in Fig.~\ref{fig:S9}c, magnetic field sweeps similar to those performed on Section D1 demonstrate the suppression of the superconducting gap, but with no switching events, as expected given that a micromagnet array is absent for section D3.\\
\\
\noindent
[1] C.W. Groth, M. Wimmer, A.R. Akhmerov and X. Waintal, New Journal of Physics \textbf{16}, 063065 (2014).\newline
[2] A.P. Higginbotham \textit{et al}, Nature Physics \textbf{11}, 1017 (2015).\newline
%[2] A.P. Higginbotham, S.M. Albrecht, G. Kir{\v s}anskas, W. Chang, F. Kuemmeth, P. Krogstrup, T.S. Jespersen, J. Nyg{\aa}rd, K. Flensberg and C.M. Marcus, Nature Physics \textbf{11}, 1017 (2015).\newline
[3] M. Tinkham, \textit{Introduction to Superconductivity} (McGraw-Hill Education, 1996).\newline
[4] W. Rave and A. Hubert, IEEE Transactions on Magnetics \textbf{36}, 3886 (2000).\newline
[5] X. Liu, J.N. Chapman, S. McVitie and C.D.W. Wilkinson, Journal of Applied Physics \textbf{96}, 5173 (2004).\newline
[6] M. Kjaergaard, K. W{\"o}lms and K. Flensberg, Physical Review B \textbf{85}, 020503 (2012).\newline
[7] S. Turcotte, S. Boutin, J.C. Lemyre, I. Garate and M. Pioro-Ladri{\`e}re, Physical Review B \textbf{102}, 125425 (2020).

% \bibliography{bibliography}
%%%%%%%%%%%%%%%%%%%

\end{document}